\documentclass[aps,prl,floats,twocolumn]{revtex4}
\usepackage{graphicx}

\input{tcilatex}

\begin{document}

\title{Manipulating the Quantum State of an Electrical Circuit}
\author{D. Vion\footnote{To whom correspondence should be addressed;
E-mail: {\tt vion@drecam.saclay.cea.fr}},
A. Aassime, A. Cottet, P. Joyez, H. Pothier, C. Urbina\footnote{Member of CNRS.}, D. Esteve, M.H. Devoret\footnote{Present address: Applied Physics,
Yale University, New Haven, CT 6520, USA}}
\affiliation{Quantronics Group, Service de Physique de l'Etat Condens\'{e}, Direction des Sciences de la Mati\`{e}re,
CEA-Saclay, \\
91191 Gif-sur-Yvette, France.\\
}
\date{Dec. 21, 2001}

\begin{abstract}
\medskip
We have designed and operated a superconducting tunnel junction circuit that
behaves as a two-level atom: the ``quantronium''. An arbitrary evolution of
its quantum state can be programmed with a series of microwave pulses, and a
projective measurement of the state can be performed by a pulsed readout
sub-circuit. The measured quality factor of quantum coherence $Q_{\varphi
}\simeq 25000$ is sufficiently high that a solid-state quantum processor
based on this type of circuit can be envisioned.
\end{abstract}

\maketitle

Can we build machines that actively exploit the fundamental properties of
quantum mechanics, such as the superposition principle or the existence of
entangled states? Applications such as the transistor or the laser, often
quoted as developments based on quantum mechanics, do not actually answer
this question. Quantum mechanics enters into these devices only at the level
of material properties but their state variables such as voltages and
currents remain classical. Proposals for true quantum machines emerged in
the last decades of the 20th century and are now being actively explored:
quantum computers \cite{QC}, quantum cryptography communication systems \cite%
{QS2} and detectors operating below the standard quantum limit \cite{QA3}.
The major difficulty facing the engineer of a quantum machine is decoherence %
\cite{decoh2}. If a degree of freedom needs to be manipulated externally, as
in the writing of information, its quantum coherence usually becomes very
fragile. Although schemes that actively fight decoherence have recently been
proposed \cite{QEC,QEC2}, they need very coherent quantum systems to start
with. The quality of coherence for a two-level system can be quantitatively
described by the quality factor of quantum coherence $Q_{\varphi }=$ $\pi
\nu _{01}T_{\varphi }$ where $\nu _{01}$ is its transition frequency and $%
T_{\varphi }$ is the coherence time of a superposition of the states. It is
generally accepted that for active decoherence compensation mechanisms, $%
Q_{\varphi }$'s larger than $10^{4}\nu _{01}\,t_{\mathrm{op}}$ are
necessary, $t_{\mathrm{op}}$ being the duration of an elementary operation %
\cite{minQ}.

Among all the practical realizations of quantum machines, those involving
integrated electrical circuits are particularly attractive. However, unlike
the electric dipoles of isolated atoms or ions, the state variables of a
circuit like voltages and currents usually undergo rapid quantum decoherence
because they are strongly coupled to an environment with a large number of
uncontrolled degrees of freedom \cite{Schoen}. Nevertheless, superconducting
tunnel junction circuits \cite{MQT2,Nakamura,Mooij,Lukens,Han} have
displayed $Q_{\varphi }$'s up to several hundred \cite{Martinis} and
temporal coherent evolution of the quantum state has been observed on the
nanosecond time scale \cite{Nakamura,Nakamura3} in the case of the single
Cooper pair box \cite{quantro}.\ We report here a new circuit built around
the Cooper pair box with $Q_{\varphi }$ in excess of $10^{4}$, whose main
feature is the separation of the\ write and readout ports \cite{design,Zorin}%
. This circuit, which behaves as a tunable artificial atom, has been
nicknamed a ``quantronium''. The basic Cooper pair box consists of a low
capacitance superconducting electrode, the ``island'', connected to a
superconducting reservoir by a Josephson tunnel junction with capacitance $%
C_{j}$ and Josephson energy $E_{J}$.\ The junction is biased by a voltage
source $U$ in series with a gate capacitance $C_{g}$. In addition to $E_{J}$
the box has a second energy scale, the Cooper pair Coulomb energy $%
E_{CP}=(2e)^{2}/2\left( C_{g}+C_{j}\right) $. When the temperature $T$ and
the superconducting gap $\Delta $ satisfy $k_{B}T\ll \Delta /\ln \mathcal{N}$
and $E_{CP}\ll \Delta $, where $\mathcal{N}$ is the total number of paired
electrons in the island, the number of excess electrons is even \cite%
{Tuominen,parity}.\ The Hamiltonian of the box is then%
\begin{equation}
\hat{H}=E_{CP}\left( \hat{N}-N_{g}\right) ^{2}-E_{J}\cos \hat{\theta}~\text{,%
}  \label{Ham}
\end{equation}%
where $N_{g}=C_{g}U/2e$ is the dimensionless gate charge and $\hat{\theta}$
the phase of the superconducting order parameter in the island, conjugate to
the number $\hat{N}$ of excess\ Cooper pairs in it \cite{quantro}.

In our experiment, $E_{J}\simeq E_{CP}$ and neither $\hat{N}$ nor $\hat{%
\theta}$ is a good quantum number. The box thus has discrete quantum states
that are quantum superpositions of several charge states with different $N$.
Because the system is sufficiently non-harmonic, the ground $\left|
0\right\rangle $ and first excited $\left| 1\right\rangle $ energy
eigenstates form a two-level system.\ This system corresponds to an
effective spin one-half $\vec{s}$ whose Zeeman energy $h\nu _{01}$ goes to a
minimal value close to\emph{\ }$E_{J}$ when $N_{g}=1/2$. At this particular
bias point both states $\left| 0\right\rangle \,$($s_{z}=+1/2$) and $\left|
1\right\rangle $ ($s_{z}=-1/2$) have the same average charge $\left\langle 
\hat{N}\right\rangle =1/2,$ and consequently the system is immune to first
order fluctuations of the gate charge. Manipulation of the quantum state is
performed by applying microwave pulses $u(t)$ with frequency $\nu \simeq \nu
_{01}$ to the gate, and any superposition $\left| \Psi \right\rangle =\alpha
\left| 0\right\rangle +\beta \left| 1\right\rangle $ can be prepared.

A novel type of readout has been implemented in this work. The single
junction of the basic Cooper pair box has been split into two nominally
identical junctions in order to form a superconducting loop (Fig. 1). The
Josephson energy $E_{J}$ in Eq. \ref{Ham} becomes $E_{J}\cos (\hat{\delta}%
/2) $ \cite{Likharev}, where $\hat{\delta}$ is an additional degree of
freedom, the superconducting phase difference across the series combination
of the two junctions \cite{Averin}. The two states are discriminated not
through the charge $\left\langle \hat{N}\right\rangle $ on the island \cite%
{Nakamura,Delsing}, but through the supercurrent in the loop $\left\langle 
\hat{I}\right\rangle =(2e/\hbar )\left\langle \partial \hat{H}/\partial \hat{%
\delta}\right\rangle $. This is achieved by entangling $\vec{s}$ with the
phase $\hat{\gamma}$ of a large Josephson junction with Josephson energy $%
E_{J0}\approx 20E_{J}$, inserted in the loop \cite{design,Buisson}. The
phases are related by $\hat{\delta}=\hat{\gamma}+$ $\phi $, where $\phi
=2e\Phi /\hbar $, $\Phi $ being the external flux imposed through the loop.
The junction is shunted by a capacitor $C$ to reduce phase fluctuations. A
trapezoidal readout pulse $I_{b}(t)$ with a peak value slightly below the
critical current $I_{0}=2eE_{J0}/\hbar $ is applied to the parallel
combination of the large junction and the small junctions (Fig. 1C). When
starting from $\langle \hat{\delta}\rangle \approx 0$, the phases $\langle 
\hat{\gamma}\rangle $ and $\langle \hat{\delta}\rangle $ grow during the
current pulse, and consequently a $\vec{s}$-dependent supercurrent develops
in the loop. This current adds to the bias-current in the large junction,
and by precisely adjusting the amplitude and duration of the $I_{b}(t)$\
pulse, the large junction switches during the pulse to a finite\emph{\ }%
voltage state with a large probability $p_{1}$ for state $\left|
1\right\rangle $ and with a small probability $p_{0}$ for state $\left|
0\right\rangle $\thinspace \cite{design}. This readout scheme is similar to
the spin readout of Ag atoms in a Stern and Gerlach apparatus, in which the
spin is entangled with the atom position. For the parameters of the
experiment, the efficiency of this projective measurement should be $\eta
=p_{1}-p_{0}=0.95$ for optimum readout conditions. The readout is also
designed so as to minimize the $\left| 1\right\rangle \rightarrow \left|
0\right\rangle $ relaxation rate using a Wheatstone-bridge-like symmetry.
Large ratios $E_{J0}/E_{J}$ and $C/C_{j}$ provide further protection from
the environment. Just as the system is immune to charge noise at $N_{g}=1/2$%
, it is immune to flux and bias current noise at $\phi =0$ and $I_{b}=0,$
where $\hat{I}=0$. The preparation of the quantum state and its manipulation
are therefore performed at this optimal working point.

\begin{figure}[ht]
\includegraphics*[width=0.4\textwidth]{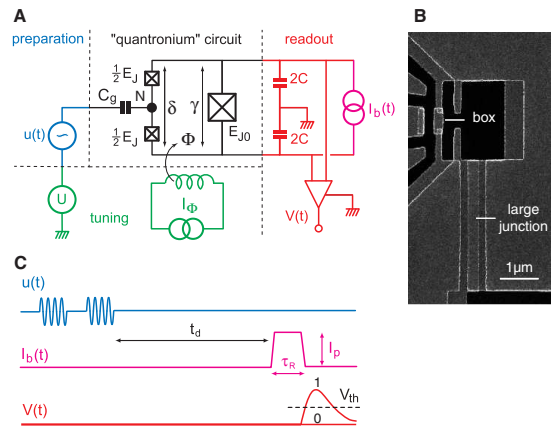}
\caption{(A) Idealized
circuit diagram of the ``quantronium'', a quantum coherent circuit with its
tuning, preparation and readout blocks. The circuit consists of a Cooper
pair box island (black node) delimited by two small Josephson junctions
(crossed boxes) in a superconducting loop. The loop also includes a third,
much larger Josephson junction shunted by a capacitance $C$.
The Josephson energies of the box and the large junction are $E_{J}$
and $E_{J0}$. The Cooper pair number $N$
and the phases $\delta$ and
$\gamma$ are the degrees of freedom of the circuit.
A dc voltage $U$ applied to the gate capacitance $C_{g}$
and a dc current $I_\phi$
applied to a coil producing a flux $\Phi$ in the circuit
loop tune the quantum energy levels. Microwave pulses $u(t)$
applied to the gate prepare arbitrary quantum states of the circuit. The
states are readout by applying a current pulse $I_{b}(t)$
to the large junction and by monitoring the voltage $V(t)$
across it. (B) Scanning electron micrograph of a sample. (C) Signals
involved in quantum state manipulation and measurement. Top: Microwave
voltage pulses $u(t)$ are applied to the gate for state
manipulation. Middle: A readout current pulse $I_{b}(t)$
with amplitude $I_{p}$ is applied to the large junction
$t_{d}$ after the last microwave pulse. Bottom: Voltage
$V(t)$ across the junction. The occurence of a pulse depends
on the occupation probabilities of the energy eigenstates. A discriminator
with threshold $V_{th}$ converts $V(t)$
into a boolean output for statistical analysis.}
\end{figure}

A quantronium sample is shown in Fig.~1B. It was fabricated with standard
e-beam lithography and aluminum evaporation. The sample was cooled down to 15~mK in a dilution refrigerator. The switching of the large
junction \cite{plasma} to the finite voltage state is detected by measuring
the voltage across it with a room temperature preamplifier followed by a
discriminator. By repeating the experiment, the switching probability, and
hence the occupation probabilities of the $\left| 0\right\rangle $ and $%
\left| 1\right\rangle $ states, can be determined.

The readout part of the circuit was tested by measuring the switching
probability $p$ at thermal equilibrium as a function of the pulse height $%
I_{p}$, for a readout pulse duration of $\tau _{r}=100~\mathrm{ns}$. The
discrimination between the estimated currents for the $\left| 0\right\rangle 
$ and $\left| 1\right\rangle $ states was found to have an efficiency of $%
\eta =0.6$, which is lower than the expected $\eta =0.95$. Measurements of
the switching probability as a function of temperature and repetition rate
indicate that the discrepancy between the theoretical and experimental
readout efficiency could be due to an incomplete thermalization of our last
filtering stage in the bias current line.

\begin{figure}[ht]
\includegraphics*[width=0.4\textwidth]{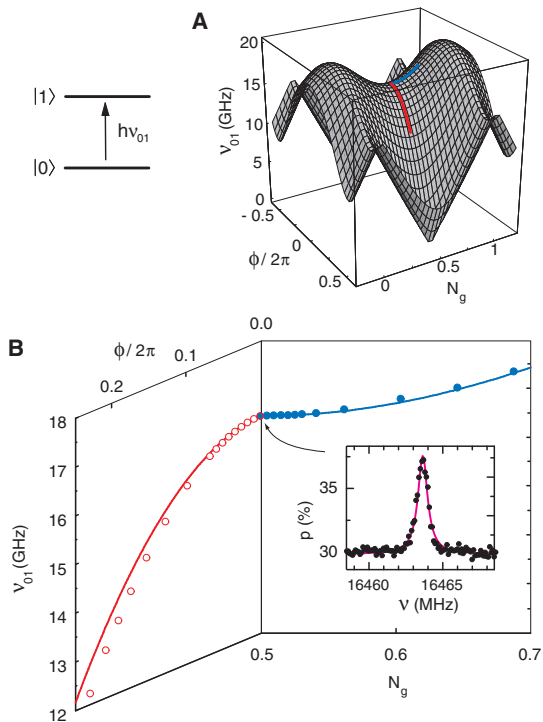}
\caption{(A) Calculated
transition frequency $\nu _{01}$ as a function of
$\phi$ and $N_{g}$ for
$E_{J}=0.865\,k_{B}K$ and $E_{J}/E_{CP}=1.27$.
The saddle point at the intersection of the blue and red
lines is an ideal working point where the transition frequency is
independent, to first order, of the bias parameters. (B) Measured center
transition frequency (symbols) as a function of reduced gate charge $N_{g}$
for reduced flux $\phi =0$
[right panel, blue line in (A)] and as a function of $\phi$
for $N_{g}=0.5$ [left panel, red line in
(A)], at 15~mK. Spectroscopy is performed by measuring
the switching probability $p$ ($10^{5}$
events) when a continuous microwave irradiation of variable frequency is
applied to the gate before readout ($t_{d}<100~ns$).
Continuous line: Theoretical best fit leading to $E_{J}$
and $E_{J}/E_{CP}$ values indicated above. Inset:
Lineshape measured at the optimal working point $=0$
and $N_{g}=0.5$ (dots). Lorentzian fit
with a FWHM $\Delta \nu _{01}=0.8~\rm{MHz}$ and a
center frequency $\nu _{01}=16463.5~\rm{MHz}$ (solid
line).}
\end{figure}

Spectroscopic measurements of $\nu _{01}$were performed by applying to the
gate a weak continuous microwave irradiation suppressed just before the
readout pulse. The variations of the switching probability as a function of
the irradiation frequency display a resonance whose center frequency evolves
with dc gate voltage and flux as the Hamiltonian predicts, reaching $\nu
_{01}\simeq 16.5~\mathrm{GHz}$ at the optimal working point (Fig.~2). The
small discrepancy between theoretical and experimental value of the
transition frequency at non-zero magnetic flux is attributed to flux
penetration in the small junctions not taken into account in the model.
These spectroscopic data have been used to precisely determine the relevant
circuit parameters, $E_{J}=0.865\,k_{B}\mathrm{K}$ and $E_{J}/E_{CP}=1.27$.
At the optimal working point, the linewidth was found to be minimal with a $%
0.8~$MHz full width at half-maximum (FWHM). When varying the delay between
the end of the irradiation and the readout pulse, the resonance peak height
decays with a time constant $T_{1}=1.8~\mathrm{\mu s}$. Supposing that the
energy relaxation of the system is only due to the bias circuitry, a
calculation similar to that in \cite{relax} predicts that $T_{1}\sim 10~%
\mathrm{%
{\mu}%
s}$ for a crude discrete element model. This result shows that no
detrimental sources of dissipation have been seriously overlooked in our
circuit design.

\begin{figure}[ht]
\includegraphics*[width=0.4\textwidth]{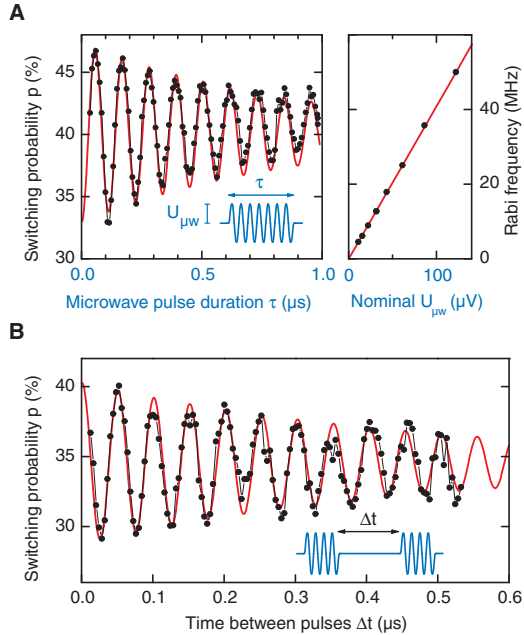}
\caption{(A) Left: Rabi
oscillations of the switching probability $p$ ($5\times 10^{4}$ events)
measured just after a resonant microwave
pulse of duration $\tau$. Data were taken at 15~mK
for a nominal pulse amplitude $U_{\mathrm{%
{\mu}%
w}}=22~%
{\mu}%
V$ (joined dots). The Rabi frequency is extracted from an
exponentially damped sinusoidal fit (continuous line). Right: Measured Rabi
frequency (dots) varies linearly with $U_{\mathrm{%
{\mu}%
w}}$, as expected. (B) Ramsey fringes of the switching
probability $p$ ($5\times 10^{4}$
events) after two phase-coherent microwave pulses separated by $\Delta t$.
Joined dots: Data at 15 \rm{mK}; the total
acquisition time was 5~mn. Continuous line: Fit by exponentially damped
sinusoid with time constant $T_{\varphi }=0.50~%
{\mu}%
s$. The oscillation corresponds to the ``beating'' of the
free evolution of the spin with the external microwave field. Its period
indeed coincides with the inverse of the detuning frequency (here
$\nu - \nu _{01}=20.6~$\rm{MHz}).}
\end{figure}

Controlled rotations of $\vec{s}$ around an axis $x$ perpendicular to the
quantization axis $z$ have been performed. Before readout, a single pulse at
the transition frequency with variable amplitude $U_{\mathrm{%
{\mu}%
w}}$ and duration $\tau $ was applied. The resulting change in switching
probability is an oscillatory function of the product $U_{\mathrm{%
{\mu}%
w}}\tau $ (Fig.~3A), which is in agreement with the theory of Rabi
oscillations \cite{Rabi}, proving that the resonance indeed arises from a
two-level system. The proportionality ratio between the Rabi period and $U_{%
\mathrm{%
{\mu}%
w}}\tau $ was used to calibrate microwave pulses for the application of
controlled rotations of $\vec{s}$.

Rabi oscillations correspond to a driven coherent evolution but do not give
direct access to the intrinsic coherence time $T_{\varphi }$ during a free
evolution of $\vec{s}$ .\ This $T_{\varphi }$ was obtained by performing a
Ramsey fringe experiment \cite{Ramsey} on which atomic clocks are based. One
applies to the gate two phase coherent microwave pulses each corresponding
to a $\pi /2$ rotation around $x$ \cite{noteangle} and separated by a delay $%
\Delta t$ during which the spin precesses freely around $z$. For a given
detuning of the microwave frequency, the observed decaying oscillations of
the switching probability as a function of $\Delta t$ (Fig.~3B) correspond
to the ``beating'' of the spin precession with the external microwave field %
\cite{noteRamsey}. The oscillation period agrees exactly with the inverse of
the detuning, allowing a measurement of the transition frequency with a
relative accuracy of $6\times 10^{-6}$. The envelope of the oscillations
yields the decoherence time $T_{\varphi }\simeq 0.50~\mathrm{%
{\mu}%
s}$. Given the transition period $1/\nu _{01}\simeq 60~\mathrm{ps}$, this
means that $\vec{s}$ can perform on average 8000 coherent free precession
turns.

In all the time domain experiments on the quantronium, the oscillation
period of the switching probability agrees closely with theory, which proves
controlled manipulation of $\vec{s}$. However, the amplitude of the
oscillations is smaller than expected by a factor of 3 to 4. This loss of
contrast is likely to be due to a relaxation of the level population during
the measurement itself.

\ In order to understand what limits the coherence time of the circuit,
measurements of the linewidth $\Delta \nu _{01}$ of the resonant peak as a
function of $U$ and $\Phi $ have been performed. The linewidth increases
linearly when departing from the optimal point $(N_{g}=1/2,~\phi
=0,~I_{b}=0) $. This dependence is well accounted for by charge and phase
noises with root mean square deviations $\Delta N_{g}=0.004$ and $%
\Delta (\delta /2\pi )=0.002$ during the time needed to record the
resonance. The residual linewidth at the optimal working point is
well-explained by the second order contribution of these noises. The
amplitude of the charge noise is in agreement with measurements of $1/f$
charge noise \cite{PTB}, and its effect could be minimized by increasing the 
$E_{J}/E_{CP}$ ratio. The amplitude of the flux noise is unusually large %
\cite{Wellstood} and should be significantly reduced by improved magnetic
shielding. An improvement of $Q_{\varphi }$ by an order of magnitude thus
seems possible. Experiments on quantum gates based on the controlled
entanglement of several capacitively coupled\emph{\ }quantronium circuits
could be already performed with the level of quantum coherence achieved in
the present experiment.

\begin{acknowledgments}
The indispensable technical work of P. Orfila is gratefully acknowledged.
This work has greatly benefited from direct inputs from J. M.\ Martinis and
Y. Nakamura. The authors acknowledge discussions with P. Delsing, G. Falci,
D. Haviland, H.\ Mooij, R.\ Schoelkopf, G.\ Sch\"{o}n and G. Wendin. Partly
supported by the European Union through contract IST-10673 SQUBIT and the
Conseil G\'{e}n\'{e}ral de l'Essonne through the EQUM project.
\end{acknowledgments}


\begin{thebibliography}{99}
\bibitem{plasma} For $C=1~\mathrm{pF}$ and $I_{0}=0.77~\mathrm{%
{\mu}%
A,}$ the bare plasma frequency of the large junction is $\omega _{p}/2\pi
\simeq 8~\mathrm{GHz}$, well below $\nu _{01}$.
\bibitem{noteangle} In practice, the rotation axis does not need to be $x$,
but the rotation angle of the two pulses is always adjusted so as to bring a
spin initially along $z$ into the plane perpendicular to $z$.
\end{thebibliography}
\end{document}